\documentclass[aps,prl,twocolumn,showpacs,groupedaddress,14pt]{revtex4}

\usepackage[english]{babel}
\usepackage{mathtext,amssymb,amsmath,float}
\usepackage{graphics}
\usepackage{epsfig}

\begin{document}

\title{Evolving dynamical networks with transient cluster activity}

%

\author{Oleg~V.~Maslennikov}
\email{olmaov@neuron.appl.sci-nnov.ru}
\author{Vladimir~I.~Nekorkin}
\affiliation{Institute of Applied Physics of RAS, Nizhny Novgorod, Russia}
\affiliation{N.I.Lobachevsky State University of Nizhny Nodgorod, Nizhny Novgorod, Russia}


\date{\today}
\begin{abstract}
We study transient sequential dynamics of evolving dynamical networks, i.e., those having active nodes and links and activity-dependent topology. We show that such networks can generate sequences of metastable cluster states where each state is a cyclic sequence of clusters following each other in a certain order. We found the way how the sequences generated by such networks can be robust against background noise, small perturbations of initial conditions, and parameter detuning, and at the same time, can be sensitive to input information.
\end{abstract}

\maketitle

\section{Introduction}
Dynamics of complex systems consisting of many interacting active elements have recently attracted much attention when studying collective phenomena in natural and social sciences~\cite{Strogatz2001,Albert2002,Newman2003,Boccaletti2006}. Most of the works in this field focus on how the network behavior is influenced by its static, i.e., constant connection topology. However, in many real networked systems the connectivity varies in time and is inherently dependent on the dynamics of nodes and links, and vice versa. For describing these complex networks that have interdependent dynamical nodes and links and evolving topology, a new general formalism called an Evolving Dynamical Network was proposed in Ref.~\cite{Gorochowski2011} by Gorochowski et al. In this framework, the complex network behavior is considered as a two-level process. First, the network topology \emph{evolves}, i.e., passes though different structure configurations according to some evolutionary operator. Second, at each structure configuration (also called a generalized dynamical graph), the \emph{dynamics} of the network, i.e. of nodes and links, takes place governed by the associated dynamical mapping (see also Ref.~\cite{Belykh2014}). Such two-level representation as the structure evolution and the network dynamics allows one to adequately describe different collective behaviors in complex active systems.

One of the important and topical problems in the complex networks theory is transient dynamics in the form of sequences of metastable states where each state is a synchronous activation of a certain group of network nodes. Such sequential behavior is typical for a wide range of neural networks~\cite{Rabinovich2008,Jones2007,Rabinovich2014} and has several interesting properties. First, a sequence of generated states is sensitive to an input stimulus, and different stimuli evoke different activity patterns (the property of selectivity). Second, the network dynamics is robust against noise and small perturbations of initial conditions and thus is reproducible (the property of structural stability). Given these features, the study of complex networks with switching dynamics is based on completely different approaches compared with attractor dynamical systems because in this case we are  interested not in a final regime to which the network asymptotically tends but in the whole sequence of states. Some models satisfying these conditions have been so far proposed in the form of static networks~\cite{Rabinovich2001,Huerta2004,Rabinovich2006,Muezzinoglu2010}. Their dynamics may be reduced to the (generalized) Lotka-Volterra equations and are based on the winnerless competition principle: a sequential activation of different metastable states wherein at every time moment only one state is active. The dynamical image of metastable states is saddle equilibria and the switching dynamics between them are formed by heteroclinic channels that connect the saddles. An informational stimulus sets a specific topology in the network reflected by a heteroclinic channel in the phase space: a set of trajectories moving from the neighborhood of one saddle to that of another and so forth. Trajectories that belong to the channel form the so-called  transient metastable dynamics of the network.

Switching dynamics are found in static oscillatory networks mimicking neural activity in the framework of different models. For example, in Ref.~\cite{Casado2003} the sequential activation and deactivation of neural groups by stimulation was demonstrated for a network of map-based neurons with inhibitory connections. For a network of the FitzHugh-Nagumo neurons, the existence of stable heteroclinic sequences and channels which connect saddle limit cycles was shown in Ref.~\cite{Komarov2013}. In Refs.~\cite{Nekorkin2010,Nekorkin2011} an alternative approach to describing switching dynamics was proposed. For a network of the Morris-Lecar neurons it was shown that metastable states can be formed by oscillatory firings which appear and disappear through dynamic bifurcations. In Refs.~\cite{Gros2007,Linkerhand2013} the authors proposed the use of attractor relict networks to describe autonomous transient dynamics in neural systems.

Switching dynamics were also studied for networks of phase oscillators with variable couplings~\cite{Ashwin2007}. For a network of 5 nodes, a heteroclinic network of metastable states was found where each state is a particular set of three phase clusters. It was explored how the noise, stimulation, and detuning influence the route of switching states in the heteroclinic network. The possibility was demonstrated~\cite{Wordsworth2008} of coding and decoding of spatiotemporal information with the use of switching dynamics in such networks.

\begin{figure}[h]
\begin{center}
  \includegraphics[width=0.48\textwidth]{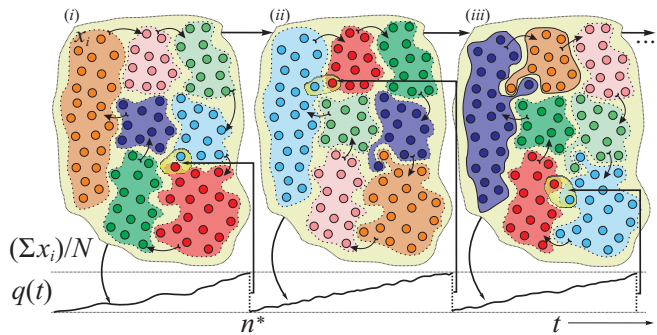}
  \caption{Switching dynamics in the evolving dynamical network. A network consists of $N$ nodes characterized by the $x_i$-dynamics ($i=1\dots N$). The collective network behavior is a sequence of cluster states [labeled by $(i)$, $(ii)$,  and $(iii)$]. Each state is characterized by a cyclic sequence of clusters (groups of nodes with the same color) which follow each other with a certain order (marked by arrows). The nodal averaged activity $N^{-1} \sum x_i$ influences  the network topology through the variable $q(t)$. The network is in a given cluster state [e.g., $(i)$] while $q(t)$ is slowly increasing. When $q$ reaches a threshold at the moment of $n^*$, the topology is rewired which results in a new cluster state $(ii)$. }
  \label{fig:scheme}
  \end{center}
\end{figure}

In the present work, we propose a new model of evolving dynamical networks which can generate sequences of metastable states formed by the clusters of synchronous activity in response to the informational input (see Fig.~\ref{fig:scheme}). We show that the dynamics of these networks are selective to stimuli and robust against noise and small perturbations of initial conditions.  Internodal interaction with inhibitory connections leads to the switching dynamics of the second level: clusters cyclically follow each others. Due to the feedback, the collective nodal dynamics finally results in rewiring of the network topology. Consequently, a new pattern, i.e., another cyclic  sequence of clusters appears. This is a mechanism for the structure evolution as well as the switching dynamics of the first level: a sequence of cluster states. Generally, this sequence is not cyclic unlike the switching dynamics of the second level, but is determined by the input stimulus. We claim that even in a small network there are a large number of possible cluster states which are connected to some heteroclinic network in the state space. Due to this  property, a variety of different sequences of cluster states can be realized in the system in response to input stimuli. In other words, the evolving dynamical networks have a high informational capacity compared to the static ones. For definiteness, in the following sections we illustrate these ideas by studying an example of the network with active nodes governed by the maps with intrinsic chaotic dynamics mimicking neural behavior.

\section{Dynamics of nodes}
Consider a network of $N$ nodes with only inhibitory connections between them which is typical for neural networks with switching dynamics. The nodal activity is described by the following map~\cite{Nekorkin2007,Courbage2007,Courbage2010}:

\begin{equation}\label{eq:map2d}
\begin{cases}
x_{i,n+1} = x_{i,n} + F_H(x_{i,n}) - y_{i,n} + I_{i,n},\\
y_{i,n+1} = y_{i,n} + \varepsilon(x_{i,n} - J_i), i=1,...,N,
\end{cases}
\end{equation}

where $n = 0,1,2,\dots$ is discrete time,  the variables $x_{i,n}$ and $y_{i,n}$ characterize the state of the $i$-th node at the moment of $n$. The nonlinear function $F_H(x) = x (x-a) (1-x) - \beta H(x-d)$, where $H(x)$ is the Heaviside step function, and the parameters $a$, $\beta$, and $d$ control the dynamical oscillatory regime. The parameter $\varepsilon$ determines the rate for the variable $y_{i}$, the parameter $J_i$ characterizes the excitatory properties of the nodes, and the term $I_{i}$ is an external influence on the $i$-th node. We fix the parameters $a=0.1$, $\beta=0.3$, $d=0.45$, and $\varepsilon=10^{-3}$. For these values, the system~\eqref{eq:map2d} has the following dynamical properties. In the absence of an input ($I_{i,n}=0$), the map~\eqref{eq:map2d} has a unique stable fixed point with the coordinates $x^*_{i}=J_i$ and $y^*_i = F_H(J_i)$. For relatively small perturbations, the system quickly returns to the equilibrium. For strong enough inhibitory input  $I_{i}<0$, a positive response is generated in the form of burst, i.e., a sequence of spikes. Note that for the parameter values chosen, the first equation in~\eqref{eq:map2d} is a Lorenz-type map and has a chaotic attractor for fixed values $y_{i}$ from some range. The variation of $y_{i}$ in time forms the relaxation dynamics of~\eqref{eq:map2d} with both regular and chaotic features (for more details, see~\cite{Courbage2007,Maslennikov2013}). Thus, the nodal response of the nodes generally displays intrinsic chaotic dynamics.

The term $I_{i,n}$ takes into account the impact on the  $i$-th node from other nodes as well as from noisy inputs. The influence of the other network nodes is described by the following equation:
\begin{equation} \label{eq:chem1}
    I_{i,n} = - g \sum\limits_{j=1, j\neq i}^{j=N} G_{ij, n} (x_{i,n}-\nu) H(x_{j,n}-\theta),
\end{equation}
where the coefficient $g$ defines the coupling strength, $\nu$ is the so-called reversal parameter, and $\theta$ is the threshold parameter. We consider only inhibitory connections in the network so we fix the values of the coupling parameters: $\nu=-0.5$ and $\theta=0.2$. The value $g=0.15$ is chosen based on the requirement that the bursting activation of one node results in the activation of the next node due to the inhibitory connection. The adjacency matrix  $G_{ij, n}$ determines the network topology: $G_{ij,n}=1$ if the $j$-th node affects the $i$-th node at the moment of $n$, and $G_{ij,n}=0$ otherwise.

\section{Cluster states}
Metastable states through which the network evolves are formed by cyclic sequences of clusters. First, we show how different regimes of $M$-cluster dynamics appear  in the network of  $N$ nodes~\eqref{eq:map2d} (where $M\leq N$). We denote the sequence of clusters that form some $M$-cluster state as follows:
$$i^{1}_{1}\dots i^{1}_{K_1} \rightarrow \dots \rightarrow i^{\alpha}_{1}\dots
i^{\alpha}_{K_{\alpha}} \rightarrow \dots \rightarrow i^{M}_{1}\dots i^{M}_{K_M}.$$

Here the first cluster comprises $K_1$ nodes with indexes $i^{1}_{1}\dots i^{1}_{K_{1}}$, i. e., during the first network configuration these nodes activate synchronously: each of them fires a burst while the remaining  nodes are at rest. Then the other clusters are fired sequentially up to the $M$-th one comprising $K_M$ nodes.  After that the cluster sequence is repeated again. For the cyclic sequence to  be generated in the network it is necessary to define a special connection topology, i.e., the adjacency matrix. For this purpose the matrix elements with indexes $(i^{\alpha}_j i^{\alpha}_k)$, where $j,k=1,\dots,K_{\alpha}$ and $\alpha=1,\dots,M$, are assumed to be zero which ensures the appearing of the $\alpha$-th cluster comprising $K_\alpha$ nodes. The matrix elements $(i^{\alpha}_j i^{\alpha+1}_k)$, where $j=1,\dots,K_{\alpha}$, $k=1,\dots,K_{\alpha+1}$, and $\alpha=1,\dots,M$ (if $\alpha=M$ then $\alpha+1$ is set equal to 1) are assumed to be equal zero in order to get switching between different clusters in the given order. The rest elements are set equal to 1 to ensure transitions from one cluster to another.

\begin{figure}[h]
\begin{center}
  \includegraphics[width=0.48\textwidth]{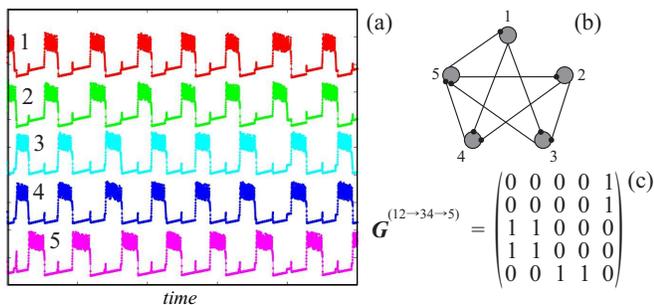}
  \caption{ (a) Cluster state in the form of cyclic sequence $12 \rightarrow 34 \rightarrow 5$ in the network of 5 nodes, (b) the corresponding network topology, and (c) the adjacency matrix  .}
  \label{fig:seq}
  \end{center}
\end{figure}

As an example we consider how  3-cluster states are formed in the network of $N=5$ nodes. Such cluster states can be of two types: either two clusters comprise two nodes and the third cluster comprises one node, or two clusters comprise one node and the third cluster comprises three nodes. The sequence  $12 \rightarrow 34 \rightarrow 5$  shown in Fig.~\ref{fig:seq}(a) is of the first type. The corresponding network topology and the adjacency matrix are shown in  Fig.~\ref{fig:seq}(b) and Fig.~\ref{fig:seq}(c). Below we illustrate general principles by the example of 5-node network and 3-cluster states with the configuration described. In such a form the model allows us to study basic mechanisms for appearing nontrivial cluster states and transitions between them remaining within the low-dimension description.

\section{Transient dynamics of cluster states}
In real networks of different nature, including neural networks, the rewiring of topology occurs permanently in accordance with nodal dynamics and external factors. However, drastic changes that have a significant impact on the network dynamics may become noticeable only with the course of time. In other words, small quantitative changes are  accumulated and at a certain moment this results in a notable qualitative rewiring of the network topology. In the simplest case, one can speak of the network evolution as a sequence of structure states following each other where each of them is characterized by a fixed network topology. Generally, the transformation rule for topology in each specific network depends on many internal and external factors. In our model we assume that the transformation law depends on the network structure and nodal dynamics and does not change over time, i.e., one can specify some evolutionary operator for the adjacency matrix.

First, we are interested in the operator that preserves the connectivity symmetry chosen, and consequently, the cluster composition for the oscillations generated. One can easily verify that the following transformation rule does not change the number and configuration of the clusters:
\begin{equation}\label{eq:P}
G_{ij,n+1} = \textbf{P}_{kl} G_{ij,n},
\end{equation}
where the operator $P_{kl}$ first changes the $k$th and $l$th rows in $G_{ij,n}$ and then in the matrix obtained changes the $k$th and $l$th columns. Which numbers $k$ and $l$ are chosen depends on the cluster state of the network at the moment of rewiring and the previous cluster state. Suppose at the moment of rewiring the nodes with indexes $i^{\alpha}_{1}\dots i^{\alpha}_{K_{\alpha}}$ are active, and before that the nodes  $i^{\beta}_{1}\dots i^{\beta}_{K_{\beta}}$ were active. There is a pair of nodes $i^{\alpha}_{k_{\alpha}}$ and $i^{\beta}_{l_{\beta}}$ from these two sets that have a minimum distance $L$ defined from nodal ordering (i.e., not a path length). The indexes of these two nodes specify the numbers $k = k_{\alpha}$ and  $l = l_{\beta}$ in Eq.~\eqref{eq:P}. To avoid ambiguity, the distance $L: i^{\alpha}_{k}\rightarrow i^{\beta}_{l}$  is calculated clockwise in a cyclically ordered set of numbered nodes $1\dots i \dots N$.

To describe the influence of the network dynamics on the structure evolution, we introduce an auxiliary variable $q$ which defines the moment $n=n^*$ of transition from one cluster state to another, controls the duration of the network sojourn at the same cluster state, and obeys the following system:
\begin{eqnarray}\label{eq:q}
q_{n+1} = q_{n} + \mu X_n, \nonumber \\
\text{if} \; q_n>1 \,  \text{then} \, q_{n+1}:=0,\nonumber \\
X_n = \frac{1}{N} \sum\limits_{i=1}^N x_{i,n}.
\end{eqnarray}

It follows from Eqs.~\eqref{eq:q} that $q$ is increasing on the average starting from zero up to the threshold $q=1$ with the rate determined by the network behavior (the mean field $X$) and a small parameter $\mu$ ($0<\mu\ll1$). The dynamics if $X$ is chaotic due to the chaotic activity of  $x_{i}$, $i=1,\dots,N$, which leads to small perturbations of $q$. After reaching the threshold $q=1$, its value is reset to zero and begins to grow according to Eq.~\eqref{eq:q}. The topology rewiring occurs at the moment of resetting of the variable $q$.

\begin{figure}[h]
\begin{center}
  \includegraphics[width=0.48\textwidth]{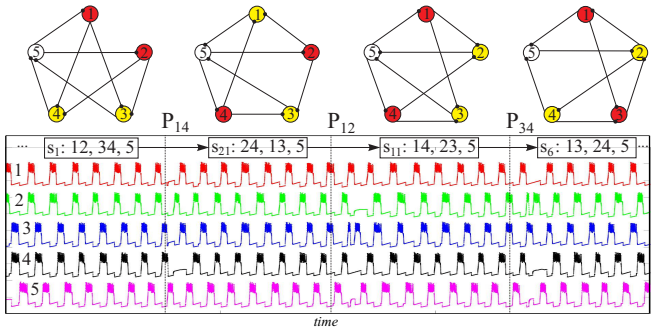}
  \caption{ Switching activity in the form of a sequence of 3-cluster states $s_{i}$ ($i=1\dots 30$) in the network of  $N=5$ nodes. Top: the time course of the network topology. The nodes of the same cluster are colored identically. Also shown are operators $P_{kl}$ which transform the adjacency matrix at the moments of rewiring indicated by vertical lines.}
  \label{fig:transient}
  \end{center}
\end{figure}

An example of how the network topology evolves over time depending on the nodal activity and how different cluster states follow each other is shown in Fig.~\ref{fig:transient} for the network of $N=5$ nodes. The network nodes colored identically belong to the same cluster. The waveform with different cluster states is shown below the structure schemes of the network.

There are 30 different 3-cluster states in the network of 5 nodes where two clusters comprise two nodes, and the third comprises one node. All of them are shown in  Table~\ref{tab:states}, where each column can be generated from another by cyclic permutation of the indexes $i_1i_2, i_3i_4, i_5$.

\begin{table}[h]
\caption{List of three-cluster states in the network of five nodes.}
\begin{center}
\begin{tiny}
\begin{tabular}{c|c|c|c|c}
\hline \hline
$s_{1} : 12,34,5$ & $s_{2} : 23,45,1$ & $s_{3} : 34,51,2$ & $s_{4} : 45,12,3$ & $s_{5} : 51,23,4$\\
$s_{6} : 13,24,5$ & $s_{7} : 24,35,1$ & $s_{8} : 35,14,2$ & $s_{9} : 14,25,3$ & $s_{10} : 25,13,4$ \\
$s_{11} : 14,23,5$ & $s_{12} : 25,34,1$ & $s_{13} : 13,45,2$ & $s_{14} : 24,15,3$ & $s_{15} : 35,12,4$ \\
$s_{16} : 23,14,5$ & $s_{17} : 34,25,1$ & $s_{18} : 45,13,2$ & $s_{19} : 15,24,3$ & $s_{20} : 12,35,4$ \\
$s_{21} : 24,13,5$ & $s_{22} : 35,24,1$ & $s_{23} : 14,35,2$ & $s_{24} : 25,14,3$ & $s_{25} : 13,25,4$ \\
$s_{26} : 34,12,5$ & $s_{27} : 45,23,1$ & $s_{28} : 51,34,2$ & $s_{29} : 12,45,3$ & $s_{30} : 23,51,4$ \\
\hline \hline
\end{tabular}
\label{tab:states}
\end{tiny}
\end{center}
\end{table}

Consider in more detail how the dynamical network evolves and different cluster states follow each other as time progresses. The first topology transformation shown in Fig.~\ref{fig:transient} results in a switching from the state $s_1: 12, 34, 5$ to the state $s_{11}: 24, 13, 5$. At the moment of switching, the nodes $3$ and $4$ are active and earlier the nodes $1$ and $2$ were active. Comparing the clockwise distances $L_i$ of the four paths $3\rightarrow1$ ($L_1=3$), $3\rightarrow2$ ($L_2=4$), $4\rightarrow1$ ($L_3=2$), and $4\rightarrow2$ ($L_4=3$), one finds that the path $4\rightarrow1$ is the shortest. So the topology changes according to the operator $P_{14}$ which results in the new state $s_{11}: 24, 13, 5$. Subsequent switchings occur according to the same algorithm. Note the system needs some time of about one or two periods to start following a new adjacency matrix with a new clustering.

\begin{figure}[h]
\begin{center}
  \includegraphics[width=0.48\textwidth]{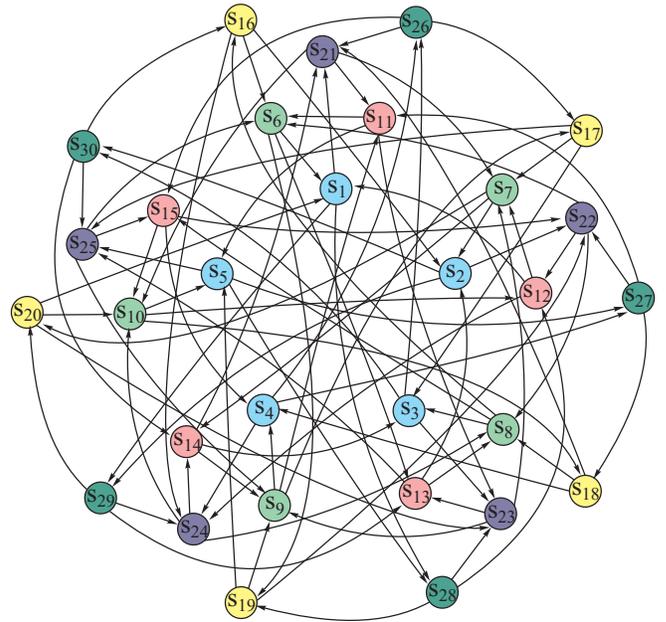}
  \caption{ Graph of 3-cluster states and transitions between them for the network of 5 nodes.}
  \label{fig:graph}
  \end{center}
\end{figure}

All the possible three-cluster states from Table~\ref{tab:states} and transients between them can be presented as a directed graph shown in Fig.~\ref{fig:graph}. Each node in this graph sends three arrows since we consider three-cluster states: the transition to one of the three possible cluster states depends on what cluster is active at the moment of rewiring.

\begin{figure}[h]
\begin{center}
  \includegraphics[width=0.48\textwidth]{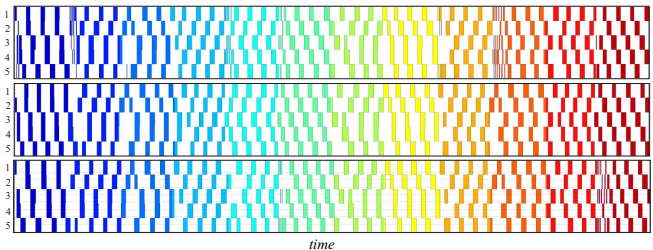}
  \caption{ The property of structural stability. There are the waveforms of switching dynamics in the network for the same input information (the initial network topology corresponds to the cluster state $s_1: 12, 34, 5$, the labeled node is $i^*=1$) for three different initial conditions, parameter detuning, and noise intensity. The same cluster states are indicated by the same colors. One can see that the sequence of cluster states $s_{1}\rightarrow s_{28}\rightarrow s_{12}\rightarrow s_{24}\rightarrow s_{14}\rightarrow s_{9}\rightarrow \dots$ is invariant and is only determined by the input.}
  \label{fig:rastr}
  \end{center}
\end{figure}

Spontaneous network dynamics and evolution described above lead to some complex route along which a trajectory moves in the space of cluster states (Fig.~\ref{fig:graph}). Generally, this route depends on initial conditions and noise because of the chaotic behavior of individual nodes. For the same initial conditions, small perturbations of the variables $x_{i}, y_{i}$ result in that the moment of rewiring can fall within activation of different clusters. To satisfy the requirement of structural stability, it is necessary for the input stimulus to contain additional information. A possible way to overcome the ambiguity when generating sequences is to label a node in the network along with specifying the initial topology. The role of this labeled node is to select the route of transition in the graph of cluster states. Namely, the topology rewiring occurs only when the cluster, containing the marker at the current moment, is active. Suppose the variable $q$ controlling the collective network state reaches the threshold, and its value is reset to zero at the moment of $n=n^*$. The network is rewired at the nearest after $n=n^*$ moment when the cluster comprising the labeled node $i^*$ is active. This labeling remains fixed until the next input comes. Such additional input information results in distinguishing a certain route in the graph of cluster states that corresponds to the given input. When noise or parameter detuning are applied, and the initial conditions for $x_{i}, y_{i}$ are perturbed, the waveforms for the same input information differ but the sequence of cluster states is invariant (see Fig.~\ref{fig:rastr}).

The activity of the evolving dynamical network is switching at two levels. First, there are transitions between different cluster states governed by the evolutionary operator. Second, each cluster state is composed by different clusters following each other cyclically according to a particular pattern.  It is clear that due to these two levels the information capacity becomes higher.  In other words, the number of information stimuli which can be transformed by the network into different spatiotemporal patterns is increased. Consider the network of 5 nodes with 3-cluster states. In the case of switchings of the second level in the \emph{static} networks there are 30 different states of this type. Therefore, the network is able to transform 30 different stimuli  (expressed by the initial topology) into robust spatiotemporal patterns. For the \emph{evolving dynamical} network of the same size, adding transitions of the second level and increasing the dimension of the input (labeled node) results in  $30\times5=150$ different sequences. If one takes the network with a greater number of nodes and considers cluster states of different types, the information capacity of the evolving dynamical network can exceeds by several orders that of the static network of the same size.

\section{Conclusion}
We have shown that evolving dynamical networks can generate sequences of metastable cluster states. Each state is a cyclic sequence of clusters following each other in a certain order. The directed graph of cluster states is formed according to a network evolutionary operator which depends on the current and previous cluster states. Transitions from one state to another in this graph happen when reaching the threshold value by the variable characterizing averaged network activity. As a result, as time progresses, the network topology is rewired, and consequently, different cluster states appear. Adding an extra input information (besides the initial topology) in the form of a labeled node provides structural stability of the sequences generated to small perturbations of initial conditions, parameter detuning, and noise, while at the same time selectivity to information stimuli. The activity-dependent topology leads to an increase of the network information capacity by orders of magnitude compared to the static network of the same size.

\section*{Acknowledgments}
This work was supported by the Russian Science Foundation (Project No. 14-12-01358).

\end{document}